\documentclass[preprint]{aastex}
\usepackage{amsmath}
\usepackage{epsfig}

\shorttitle{Neutron Star Binary Gamma-Ray Burst}
\shortauthors{Salmonson, Wilson, Mathews}

\begin{document}

\title{Gamma-Ray Bursts via the Neutrino Emission from Heated Neutron Stars}

\author{Jay D. Salmonson\altaffilmark{1} and James R. Wilson}
\affil{Lawrence Livermore National Laboratory, Livermore, CA 94550}

\author{Grant J. Mathews}
\affil{University of Notre Dame, Notre Dame, IN 46556}

\altaffiltext{1}{e-mail: salmonson@llnl.gov}

\begin{abstract}
A model is proposed for gamma-ray bursts based upon a neutrino burst
of $\sim 10^{52}$ ergs lasting a few seconds above a heated collapsing
neutron star.  This type of thermal neutrino burst is suggested by
relativistic hydrodynamic studies of the compression, heating, and
collapse of close binary neutron stars as they approach their last
stable orbit, but may arise from other sources as well.  We present a
spherically symmetric hydrodynamic simulation of the formation and
evolution of the pair plasma associated with such a neutrino burst.
This pair plasma leads to the production of $\sim 10^{51} - 10^{52}$
ergs in $\gamma$-rays with spectral and temporal properties consistent
with many observed gamma-ray bursts.
\end{abstract}

\keywords{binaries: close --- gamma rays: bursts --- gamma rays: theory --- stars: neutron}

\section{Introduction}

Understanding the origin of gamma-ray bursts (GRBs) has been a perplexing
problem since they were first  detected almost three decades
ago \citep{kso73}.  The fact that GRBs are 
distributed isotropically \citep{mfw+92} suggests a cosmological origin.
Furthermore, arcminute burst locations from BeppoSax have 
revealed that at least some $\gamma$-ray bursts
involve weak X-ray, optical, or radio transients, and are of
cosmological origin \citep{1997IAUC.6584}.  The Mg I absorption and [O II]
emission lines along the line of sight from the GRB970508 optical
transient, for example, indicate a redshift $Z \ge 0.835$
\citep{1997IAUC.6655}.  The implied distance means that this burst must have
released of order $_>^\sim 10^{51}$ ergs in $\gamma$-rays on a time
scale $\sim$ seconds.  This energy requirement has been rendered even more
demanding by other events such as GRB971214 \citep{kfw+98} which
appears to be centered on a galaxy at redshift 3.42.  This implies
that the energy of a $4\pi$ burst would have to be as much as $ 3
\times 10^{53}$ ergs, comparable to the visible light output of $\sim
10^9$ galaxies.

 Based upon the accumulated
evidence one can can now conclude that the following  four
features probably characterize the source environment:  1) 
If the total burst energies are in the
range of $10^{51}-10^{52}$ ergs, then a beaming factor of 10 to 100
is necessary;  2) The multiple peak
temporal structure of most bursts probably requires 
either multiple colliding
shocks \citep{rm94,kps98} or a single shock impinging
upon a clumpy interstellar medium \citep{mr93b,dm99};  
3) The observed afterglows imply some surrounding
material on a scale of light hours; and 4) the presence of [O II] emission
lines suggests that the bursts occur in a young, 
metal-enriched stellar population.

Some proposed sources for the production of GRBs include accretion
onto supermassive black holes, AGN's, relativistic stellar collisions,
hypernovae, and binary neutron star coalescence.  Each of these
possibilities, however, remain speculative until realistic models can
be constructed for their evolution.  In this paper we construct a
model for GRBs  produced by energetic neutrino emission from
a heated neutron star.  Our specific model for the emission
derives from the relativistic compression and heating of neutron
stars near their last stable orbit, however any scenario by which
energetic neutrino emission above a neutron star can endure for several
seconds (e.g. tidal heating, MHD induced heating, accretion shocks, 
etc) might also power the gamma-ray burst paradigm described herein.  

Our model is as follows.  A compressionally heated neutron star emits
thermal neutrino pairs which, in turn, annihilate to produce a hot
electron-positron pair plasma.  We model the expansion of the plasma
with a spherically symmetric relativistic hydrodynamics computer
program.  This simplification is justified at this stage of the
calculations since the rotational velocity of the stars is about one
third of the sound speed in the $e^+e^-$ pair plasma.  We then analyze
and compare the contributions of photons from $e^+e^-$ pair
annihilation as well as from an external synchrotron shock as the
plasma plows into the interstellar medium.  We show that the
characteristic features of GRBs, i.e. total energy, duration and
gamma-ray spectrum, can be accounted for in the context of this model.

\section{ Compression in Close Neutron Star Binaries }

It has been speculated for some time that inspiraling neutron stars
could provide a power source for cosmological gamma-ray bursts.
However, previous Newtonian and post Newtonian studies
\citep{jr96,rj98,ruffert99} of the final merger of two neutron stars
have found that the neutrino emission time scales are so short that it
would be difficult to drive a gamma-ray burst from this source.  It is
clear that a mechanism is required for extending the duration of
energetic neutrino emission.  A number of possibilities could be
envisioned, for example, neutrino emission powered by accretion
shocks, MHD or tidal interactions between the neutron stars, etc.  The
present study, however, has been primarily motivated by numerical
studies of the strong field relativistic hydrodynamics of close
neutron-star binaries in three spatial dimensions.  These studies
\citep{wm95, wmm96, mw97, mmw98a, mw99} suggest that neutron stars in
a close binary can experience relativistic compression and heating
over a period of seconds.  During the compression phase released
gravitational binding energy can be converted into internal energy.
Subsequently, up to $10^{53}$ ergs in thermally produced neutrinos can
be emitted before the stars collapse \citep{mw97,mw99}.  Here we
briefly summarize the physical basis of this model and numerically
explore its consequences for the development of an $e^+e^-$ plasma and
associated GRB.

In \citep{mw97,mw99} properties of equal-mass neutron-star binaries were
computed as a function of mass and EOS (Equation of State).  From
these studies it was deduced that compression, heating and collapse
could occur at times from a few seconds to tens of seconds before
binary merger.  Our calculation of the rates of released binding
energy and neutron star cooling suggests that interior temperatures as
hot as 70 MeV are achieved.  This leads to a high neutrino luminosity
which peaks at $L_\nu \sim 10^{53}$ ergs sec$^{-1}$.  This much
neutrino luminosity would partially convert to an $e^+e^-$ pair plasma
above the stars as is also observed above the nascent neutron star in
supernova simulations \citep{wm93}.

We should point out, however, that many papers have been published
claiming the compression is nonexistent.  In \citet{mmw98a} we
presented a rebuttal to the critics. Subsequently, however,
\citet{flan99} pointed out a spurious term in our formula for the
momentum constraint. We \citep{mw99} have corrected the momentum
constraint equation and redone a sequence of calculations for a binary
neutron star system with various angular momenta.  A compression
effect still exists which is able to release $10^{52}$ - $10^{53}$
ergs of gravitational binding energy. The compression does not occur
for corotating stars with a polytropic equation of state.  For
irrotational binary stars our compression effect is consistent
with the results of at least two other groups \citep{bonazzola99a,
bonazzola99b, mmw99, uryu99} using different numerical methods to
compute the relativistic hydrostatic equilibrium \citep{bonazzola97}.
However, these calculations were done with a polytropic equation of
state and found only very small compression; much less than 1\%.  For
the polytropic equation of state \citet{mw99} also found a compression
less than 1\%.  In simulations with the realistic, somewhat soft, EOS 
described below we
found clear evidence \citep{mw99} that significant compression, heating
and collapse still occurs for sufficiently close orbits. The reason for
this EOS dependence is straightforward.  Table \ref{eostable} shows the
realistic EOS used in the present work and the
\citep{mw99} studies.  The key difference between the polytropic and
realistic EOSs is that the adiabatic index $\Gamma$ is not constant
but  decreases at low density for a realistic EOS.
This causes the outer regions of the star to be more compact, and
therefore, less affected by tidal stabilization than for polytropes.
At the same time, the maximum central density tends to be larger for
a neutron star of a fixed baryon mass.  Therefore the relativistic
effects are more dramatic when a realistic EOS is employed.

The hydrodynamic calculations that demonstrate
compression have been made with the stars constrained to remain at 
zero temperature (i.e. efficient radiators). 
As the compression rate increases, however, it is expected
that the rate of released binding energy will exceed the
ability of the star to radiate and internal heating will result.  Large
scale off-center vortices are observed to form \citep{mmw98a}
within the stars with a characteristic
circulation time scale of $\sim 0.005$ sec.  
The maximum velocities are nearly sonic.  
Among other things, this circulatory motion
should help dissipate the compressional motion into thermal energy by shocks
thereby heating the interior of the stars.

We have run several sets of calculations with realistic neutron-star
equations of state.  We first considered stars like our earlier bench
mark cases \citep{mmw98a} with a baryon mass of 1.548 $M_\odot$
corresponding to a typical (cf. Appendix A)
gravitational mass of  M$_G$ = 1.39 M$_\odot$ and a central
density of  $\rho_c = 1.34 \times 10^{15}$ g cm$^{-3}$ in isolation.
These stars are based upon the  ``realistic'' EOS of table
\ref{eostable}
for which the maximum critical mass is $M_c = 1.575$ M$_\odot$.  
As summarized in Appendix A, this maximum mass is typical of the
somewhat soft EOS's in which relativistic particles and/or
 condensates have been included.
Parameters for this EOS were motivated by the necessity of such 
a soft EOS to obtain the correct neutrino signal in simulations 
of SN 1987A (Wilson \& Mayle 1993).
As noted in Table 5 of the Appendix this maximum mass is consistent with
the measured masses of all binary pulsar systems for which the
orbits have been well determined.

As noted above, the stars calculated using this realistic EOS 
show significant compression and released binding energy before
inspiral but do not individually collapse.  The released gravitational
energy from this calculation is summarized in Table
\ref{tableheat}. Even without the collapse instability enough internal
heating occurs to produce a significant gamma-ray burst.

We also found  \citep{mw99}  that the
individual collapse of stars would occur if the stars 
are increased in mass
from M$_G = 1.39$ to 1.44 M$_\odot$ (M$_B = 1.61$ M$_\odot$) for this
EOS. Collapse of this star system is observed to occur for very
close separation ($d = 2.4 R$) near (but before) the final stable orbit.
Thus, collapse is a reasonable possibility for typical
masses and a moderately soft EOS.  For example, collapse
would always occur prior to inspiral for
stars in the typically  observed mass range modeled with the EOS of \citet{bb95}.
For a critical mass of 1.54, even stars of initial mass of 1.35
collapse before reaching the innermost stable orbit.

Based upon the above results, we model the thermal energy deposition
due to neutron star compression as follows: we expect that the fluid
motion within the stars will quickly convert released gravitational
binding energy into thermal energy in the interior of the stars.
Thus, we estimate that the rate of thermal energy deposition is
comparable to the rate of released binding energy due to compression.
The amount of released binding energy scales with the orbital four
velocity \citep{mw97,mw99}.  An estimate of the rate of increase of
the orbital four velocity can be obtained \citep{mw97} from the
gravitational radiation timescale.  Then, from the relation between
released binding energy and increasing four velocity
\citep{mw97,mw99}, the energy deposition rate into the stars can be
deduced in approximate analytic form \citep{mw97}.  We write,
\begin{equation}
\dot E_{th}   = \frac{ (32/5) (M f)^{5/3} f E^0_{th} }
{ [1 - (64/5) (M f)^{5/3} f t]^{3/2}}~~,
\label{jay:E:energydot}
\end{equation}
where  $f$ is the orbital angular frequency  and $E_{th}^0$ is the total 
thermal energy deposited into the stars.  
In the hydrodynamic pair plasma discussions below
we consider a range of
deposited thermal energy of $E^0_{th} = 10^{51}$, $10^{52}$, $10^{53}$
ergs, consistent with the hydrodynamics simulations. 
We use the convention that
$t < 0$ and $t = 0$ is the end of energy deposition when the
neutron stars either have collapsed into two black holes or have reached 
the last stable orbit and collapsed to a single black hole.  
At the time that a typical neutron star binary system is near the last stable
obit, the orbital frequency is $\sim$ a few$\times 10^3$ sec$^{-1}$.
Hence, by Equation (\ref{jay:E:energydot}), the energy deposition rate 
would be 
\begin{equation}
\dot E_{th}  \approx 10^2 \times E^0_{th} ~~{\rm erg~sec^{-1}}~~.
\end{equation}
Thus, for $E^0_{th} = 2 \times 10^{52}$ ergs, $\dot E_{th} \approx  2 \times 10^{54}$
ergs sec$^{-1}$.

The magnitude of the neutrino
   luminosity is very critical since the subsequent fireball is formed
   by neutrino-antineutrino annihilation.  In order
to model the thermal energy emitted by neutrinos before either
   stellar or orbital collapse we have constructed a computer model
   which treats the diffusion of energy in a static neutron star.
This energy transport occurs  by a combination
   of neutrino diffusion 
 plus energy diffusion
   via a convective velocity-dependent diffusion coefficient. 
For the neutrino energy diffusion we write:
\begin{equation}
{d E_\nu \over dt} = \vec \nabla \cdot D_\nu  \vec \nabla E_\nu
\end{equation}
where the coefficient for neutrino diffusion is just the form,
\begin{equation}
D_\nu^{rad} = {c \over 3 \rho \kappa_\nu } + {R V_c \over 3}~~,
\end{equation}
where a simple estimate for the neutrino opacity $\kappa_\nu$ is used,
$\kappa_\nu \approx 9 \times 10^{43} T_{MeV}^2$ cm$^{2}$ g$^{-1}$
based upon the cross section for neutrino nuclear absorption and
scattering.  Characteristic convective velocities $V_c$ were deduced
from simulations using our three-dimensional binary neutron star code
\citep{mw99}.  We calculated an angle averaged radial
component of the fluid velocity in the frame of the star,
\begin{equation}
V_c = { 1 \over 4 \pi} \int \vert V_r \vert d(\cos \theta) d\phi ~~.
\end{equation}
For our studies, these velocities were fit with an ansatz of the form
\begin{equation}
V_c \equiv {32 \over 105} V_{c,ave} {r \over R} \sqrt{1 - r/R} ~~,
\end{equation}
where $r$ is the radial position inside a star of radius $R$ and
$V_{c,ave}$ is the volume averaged $V_c$.  This gives a good fit to
the numerical results and has the correct form in that the velocity
goes to zero at the surface and also at $r = 0$.

\begin{figure}[tb]
\centering \epsfig{figure=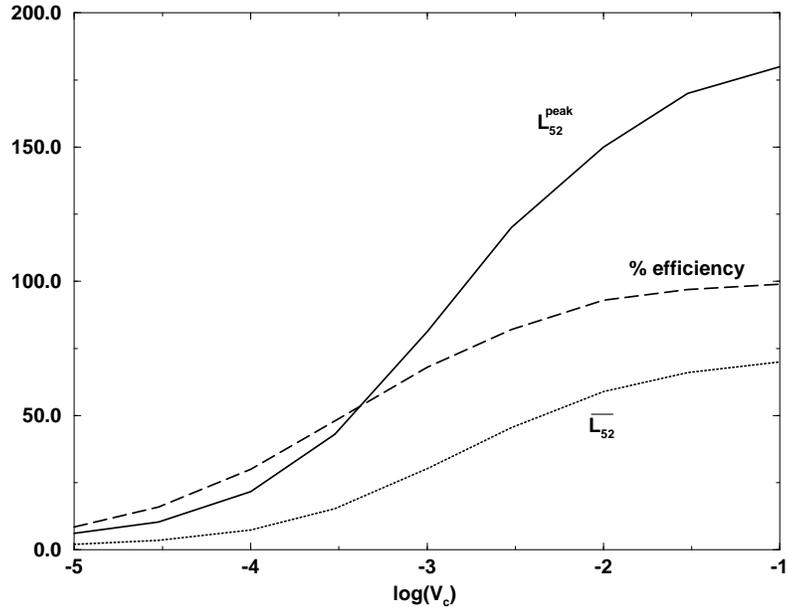, width = 9cm, angle =-90}
\caption{Luminosity from a compressed, heated neutron star as a
function of average convective velocity within the star.  Convection
substantially improves the efficiency of transport of energy to the
surface. \label{efig}} 
\end{figure}

 Energy was deposited in accordance with equation
\ref{jay:E:energydot} and the calculations terminated at time=0.  In
Figure \ref{efig} the fraction of energy released, the peak
luminosity, and $\bar L$ the average luminosity weighted by $L^{5/4}$
(see next section) are shown. The Energy input was $2\times 10^{52}$
ergs and the orbital frequency was 4000 rad sec$^{-1}$
\citep[see][]{mw99}. The average convective velocity was found to be
$\approx 0.003~c$ by analyzing the hydrodynamical calculations
\citet{mw99} of neutron star binaries. From Figure \ref{efig} we see
that a high emission efficiency and luminosity are obtained from this
convective velocity.  These produced lower thermal energies but about
the same fraction of the energy emitted, $\approx 68\%$, but the $\bar
L$ is reduced. For $E^0_{th} =0.5$ ($1.0$) $\times 10^{52}$ ergs $\bar
L= 0.75$ ($1.5$) $\times10^{53}$ ergs sec$^{-1}$.  From these
calculations we estimate that the conversion of compressional energy
to fireball energy is probably $^>_\sim 20\%$.

\section{ Neutrino Annihilation and Pair Creation  } \label{neutrinoannihilation}

\begin{figure}[tb]
\centering \epsfig{figure=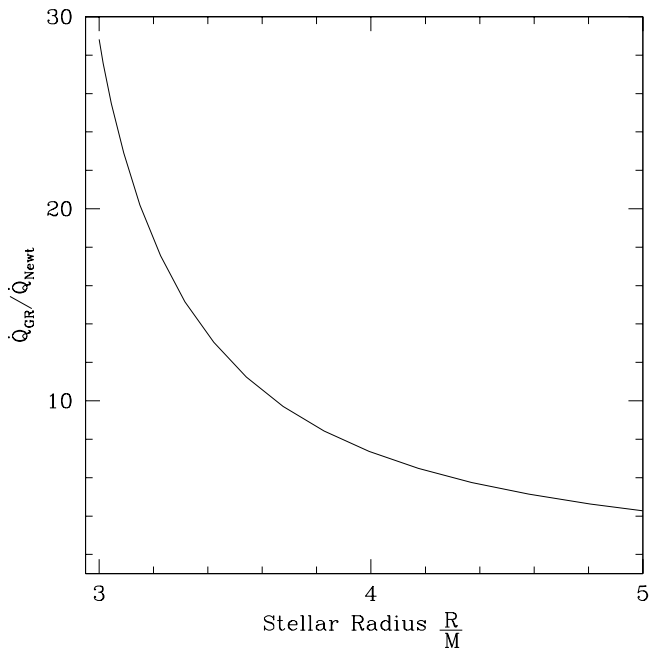, width=9cm}
\caption{General relativistic neutrino heating
augmentation ${\mathcal{F}} \equiv \dot{Q}_{GR}/\dot{Q}_{Newt}$ as a
function of neutron star neutrinosphere radius down to $R = 3M$, where
general relativistic energy deposition is $\dot{Q}_{GR}$, and
Newtonian energy deposition is $\dot{Q}_{Newt}$. \label{plotQ}}
\end{figure}

      In the previous section we have outlined a mechanism by which
neutrino luminosities of $\sim 10^{53}$ erg sec$^{-1}$ may arise from
binary neutron stars approaching their final orbits.  Here we argue
that the efficiency for converting these neutrinos into pair plasma is
probably quite high.  Neutrinos emerging from the stars will deposit
energy outside the stars predominantly by $\nu\overline{\nu}$
annihilation to form electron pairs. A secondary mechanism for energy
deposition is the scattering of neutrinos from the $e^+e^-$ pairs.
Strong gravitational fields near the stars will bend the neutrino
trajectories.  This greatly enhances the annihilation and scattering
rates \citep{sw99}. Figure \ref{plotQ} taken from \citet{sw99} shows
the relativistic enhancement factor, ${\mathcal{F}}(R/M)$, of the rate
of annihilation by gravitational bending versus the radius to mass
ratio (in units $G=c=1$).  For our employed neutron-star equations of
state the radius to mass ratio is typically between $R/M \sim 3$ and 4
just before stellar collapse.  Thus, the enhancement factor ranges
from $\sim$ 8 to 28.  Defining the efficiency of energy deposition as
the ratio of energy deposition to neutrino luminosity, then from
Equation 24 of \citet{sw99} we obtain,
\begin{equation}
        \frac{ \dot{Q} }{ L_\nu } \approx 0.03 {\mathcal{F}}(R/M)
L_{53}^{5/4}~~.
\end{equation}
   Thus, the efficiency of annihilation ranges from $\approx$0.1 to $ 0.84
\times L_{53}^{5/4}$.  For the upper range of luminosity the efficiency is
quite large.  

To better analyze the annihilation process we have
adapted the Mayle-Wilson \citep{wm93} supernova model to this problem.
We emphasize that the Mayle-Wilson model is fully general
relativistic.  To investigate this problem, a hot neutron star of the
appropriate $R/M$ was constructed and the internal temperature
adjusted to achieve the correct neutrino luminosities.  The Courant
condition requires that the time steps be quite small ($\sim 10^{-9}$
second), and zonal masses as low as $10^{-13} M_\odot$ are required    
just outside of the neutrinosphere.  Hence, the calculations could
only be evolved for a short time.  The entropy per baryon $s/k$ of the
$e^+e^-$ pair plasma is the
critical quantity for gamma-ray production.   It can be written,
\begin{equation}
s/k = \frac{4 m_b c^2 (a e^3)^{1/4} }{ 3 k \rho }~~,
\label{entropyperbaryon}
\end{equation}
where $\rho$ is the baryon density and $e$ is the total energy
density.  The entropy per baryon was found to be in the range of $10^5
- 10^6$ for the high luminosities.  For a luminosity of $10^{53}$ ergs
sec$^{-1}$, an efficiency of energy transfer from the neutrinos to the
$e^+e^-$ pair plasma due to annihilation and electron scattering was
found to be about 50 \%.  This efficiency of neutrino annihilation
determines the total energy of the pair plasma and the entropy.  This
provides the initial conditions for the subsequent fireball expansion.

\section{Pair Plasma Expansion}

Having determined the initial conditions of the hot $e^+e^-$ pair
plasma near the surface of a neutron star, we wish to follow its
evolution and characterize the observable gamma-ray emission.  To
study this we have developed a spherically symmetric, general
relativistic hydrodynamic computer code to track the flow of baryons,
$e^+e^-$ pairs, and photons.  For the present discussion we consider
the plasma deposited at the surface of a $1.45 M_\odot$ neutron star
with a radius of 10 km.

The fluid is modeled by evolving the following spherically
symmetric general relativistic hydrodynamic
equations:

\begin{equation}
\frac{\partial D }{ \partial t} = - \frac{\alpha }{ r^2} 
\frac{ \partial }{ \partial r} (\frac{r^2 }{ \alpha} D V^r)
+ \dot D_{in}
\label{jay:E:ddiff}
\end{equation}

\begin{equation}
\frac{\partial E }{ \partial t} = - \frac{\alpha }{ r^2} \frac{ \partial }{
\partial r} (\frac{r^2 }{ \alpha} E V^r) - P \biggl[ \frac{\partial W }{
\partial t} + \frac{\alpha }{ r^2} \frac{\partial }{ \partial r} (\frac{ r^2
}{ \alpha} W V^r) \biggr]
 + \dot E_{in}
\label{jay:E:ediff}
\end{equation}

\begin{equation}
\frac{\partial S_r }{ \partial t} = - \frac{\alpha }{ r^2} \frac{ \partial }
{ \partial r} (\frac{r^2 }{ \alpha} S_r V^r) - \alpha \frac{\partial P }
{ \partial r} - \alpha \frac{M }{ r^2} \biggl(\frac{D + \Gamma E }{ W}
\biggr) \biggl[ \biggl(\frac{W }{ \alpha} \biggr)^2 + \frac{(U^r)^2 }{
\alpha^4} \biggr] 
\label{jay:E:sdiff}
\end{equation}
where $D = \rho W$ and $E = \epsilon \rho W$
 are the Lorentz contracted coordinate densities of 
baryonic and thermal mass
energy ($e^+e^-$ and photons) respectively.  The quantities $\dot D_{in}$ and
$\dot E_{in}$ refer to the injected plasma from neutrino pair annihilation, and
$S_r$ is the radial
coordinate momentum density. $U_r$  is the radial component of the
covariant 4-velocity. $W\equiv \alpha U^t$ is
the generalized Lorentz factor,  $V^r$ is  the radial
coordinate three velocity, and  $\Gamma$ is an
equation of state index.
These quantities are defined by
\begin{eqnarray}
\alpha \equiv \sqrt{1 - \frac{2 M }{ r}} \quad ; \quad U_r & \equiv &
\frac{S_r }{ D + \Gamma E} \quad ; \quad W \equiv \sqrt{ 1 + U^r U_r}
\nonumber \\ & & \\ V^r \equiv \frac{U^r }{ W} \quad & ; & \quad \Gamma
\equiv 1 + \frac{ P W }{ E} \nonumber
\label{jay:E:defu}
\end{eqnarray}
To evolve the $e^+e^-$ pair plasma,  we define a pair equation.  The observed
pair annihilation rate must be corrected for relativistic effects;
specifically, time dilation slows the apparent pair
annihilation process for a fast moving fluid with respect to an observer.
  Thus, we construct
a continuity equation analogous to Equation (\ref{jay:E:ddiff}) and add
a term to account for annihilation and pair-production reactions:
\begin{equation}
\frac{\partial N_{pairs} }{ \partial t} = - \frac{\alpha }{ r^2} \frac{
\partial }{ \partial r} (\frac{r^2 }{ \alpha} N_{pairs} V^r) +
\overline{\sigma v} ((N_{pairs}^0 (T))^2 - N_{pairs}^2)/W^2 ~~.
\label{jay:E:ndiff}
\end{equation}
Here, $N_{pairs}$ is the coordinate pair number density, and
$\overline{\sigma v}$ is the Maxwellian averaged mean pair
annihilation rate per particle.  Although $\overline{\sigma v}$
depends on $T$, it varies little in the temperature range of interest,
and thus, can be taken as constant: $\overline{\sigma v} = 2.5 \times
10^{-25}$ cm$^3$ sec$^{-1}$.  $N_{pairs}^0 (T) = n_{pairs}^0 (T) W$,
where $n_{pairs}^0 (T)$ is the local proper equilibrium $e^+e^-$ pair
density at temperature T given by the appropriate Fermi integral with
a chemical potential of zero.  Zero chemical potential is a good
approximation when $N^0_{pairs}(T)$ of Equation (\ref{jay:E:ndiff}) is
important.

The total proper energy equation, including photons and $e^+e^-$ pairs
(baryon thermal energy is negligible), is
\begin{equation}
e_{tot} = a T^4 + e_{pairs}
\end{equation}
where coordinate energy in Equation (\ref{jay:E:ediff}) is related to
proper energy by $E = e_{tot} W$ and $e_{pairs}$ is the
appropriate zero chemical potential Fermi integral normalized to give
the proper $e^+e^-$ pair density $n_{pair} = N_{pairs}/W$ as
determined by Equation (\ref{jay:E:ndiff}).

The entropy per baryon (Equation \ref{entropyperbaryon}) of the wind
is crucial to the behavior of the burst.  An entropy that is too high
will create a burst which is much hotter than those observed, while an
entropy that is too low will extinguish the burst with baryons.  We
find that entropies of the order $10^7$ to $10^8$ are ideal for
producing an isotropic burst directly from the expanding pair-photon plasma.
In the calculations shown below
(Sections \ref{analysespecltcrv} \& \ref{resultspplsma}) we cover a
range of possible entropies per baryon from $10^6$ to $10^8$.  Other
possible sources of high entropy-per-baryon plasmas include the
formation of magnetized black holes \citep{rswx98,rswx99} and the
high-energy collisions ($\gamma \approx 2$) of stars in collapsing
globular clusters, which we are studying in a separate work.

We will deal with two paradigms for $\gamma-$ray production.
First, we treat the high entropy case ($s/k > 10^7$)
where the emission is  from the fireball.  Secondly, we present a
low entropy case in which gamma emission arises from the collision of
the fireball with the local interstellar medium.

In the first case, the hydrodynamic equations are evolved as the
plasma expands.  Once the system becomes transparent to Thomson
scattering, ($\int N_{pair}(r) \sigma_T dr ^<_\sim 1$ where $\sigma_T$
is the Thomson cross-section) we assume the photons are
free-streaming, the calculation is stopped and the photon gas is
analyzed to determine the photon signal.

\section{ Analysis of the Spectrum and Light curve } \label{analysespecltcrv}

We find that the photons and $e^+e^-$ pairs appear to decouple at
virtually the same time throughout the entire photon-$e^+e^-$ pair
plasma (when the cloud has reached a radius $\sim 10^{12} - 10^{13}$
cm and the temperature is typically a few 10's of eV). As such, the
photons will be well approximated as thermal and so we neglect any
radiation transport effects.  Thus, we take decoupling to be
instantaneous and to occur when the plasma becomes optically thin to
Thomson scattering.  Furthermore, we find that virtually none of the
energy deposited in the $e^+e^-$ pair plasma remains in the pairs
($\sim .001$\%). Thus, the conversion of $e^+e^-$ pair energy to
photons and baryons is very efficient.  From this simulation we derive
two observables, the time integrated energy spectrum $N(\epsilon)$ and
the total energy received as a function of observer time
$\varepsilon(t)$.

\subsection{The Spectrum}

As mentioned above, we assume that the $e^+e^-$
pairs and photons are equilibrated to the same $T$ when they decouple.
Thus, the photons in the fluid frame (denoted with a prime: $'$) make up a
Planck distribution of the form
\begin{equation}
{u'}_{\epsilon'}(T') \approx \frac{ {\epsilon'}^3 }{ exp({\epsilon' / T'}) - 1}
~~,
\end{equation}
but ${u_\epsilon / \epsilon^3}$ is a relativistic
invariant \citep{rl75}.  This implies $\epsilon / T$ is
also a relativistic invariant.  So a Planck distribution in an
emitter's rest-frame with temperature $T'$ will appear Planckian to a
moving observer, but with boosted temperature $T = T'/(\gamma (1 - v \cos
\theta))$ where $v \cos \theta$ is the component of fluid velocity (c=1)
directed toward the observer.  Thus,
\begin{equation}
u_\epsilon (\theta,v,T') \approx \frac{ \epsilon^3 }{ exp(\gamma (1 - v \cos \theta) \frac{ \epsilon }{ T'}) - 1 }
\label{jay:E:planck}
\end{equation}
gives the observed spectrum of a blackbody with rest-frame temperature
$T'$ moving at velocity $v$ and angle $\theta$ with respect to the
observer.

\begin{figure}[tb]
\centering \epsfig{figure=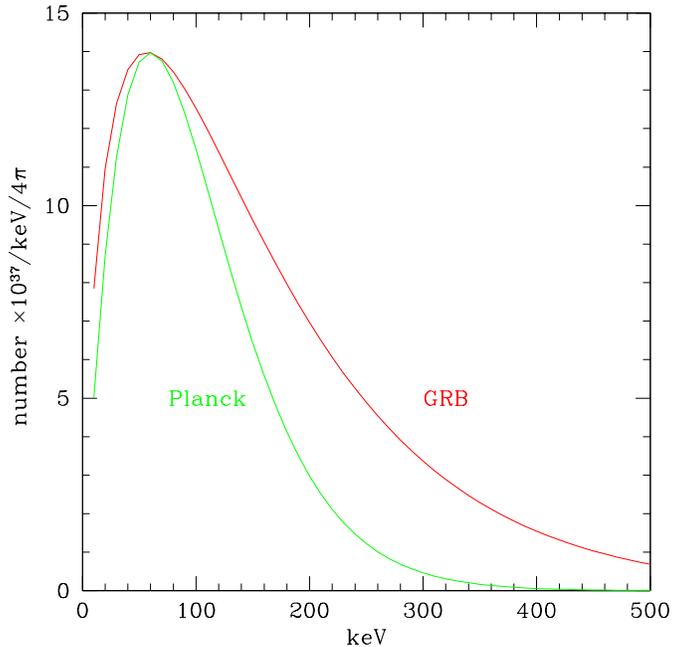, width=9cm}
\caption
{ A spectrum of a relativistically expanding spherical fireball.  A
Planck spectrum is shown for reference to show that the gamma-ray
burst spectrum is not a black-body out to several 100
keV. \label{fireballspec}}
\end{figure}

In the present case we wish to calculate the spectrum from a
spherical, relativistically expanding shell as seen by a distant
observer.  Since we know $v$, $T'$ and the radius $R$ of the shell, we
integrate over volume (i.e., shell, angle) with respect to
the observer.  We thus obtain the observed photon energy spectrum $N_\epsilon =
{\int (u_\epsilon / \epsilon)\ d^3x}$, from
 a relativistically expanding spherical
shell with radius $R$, thickness $dR$, velocity $v$, Lorentz factor
$\gamma$ and fluid-frame temperature $T'$,
 to be (in photons/eV/steradian)
\begin{equation}
N_\epsilon(v,T',R) = (5.23 \times 10^{11}) 4\pi R^2 dR \frac{\epsilon T' }{ v
\gamma} \log \Biggl[ \frac{1 - exp[- \gamma \epsilon (1 + v)/T' ] }{ 1 -
exp[ - \gamma \epsilon (1 - v)/T' ] } \Biggr]~~({\rm eV^{-1}} sr^{-1}),
\end{equation}
where $R$ is in cm.  Note, that this spectrum has a maximum at
$\epsilon_{max} \cong 1.39 \gamma T'\ eV$ for $\gamma \gg 1$.  We may
then sum this spectrum over all shells (the zones in our computer
code) of the fireball to get the total spectrum.  Figure
\ref{fireballspec} shows an example of such a spectrum up to 500 keV.
Since we assume {\it a priori} that the photons are thermal, our
spectrum has a high frequency exponential tail, but the resultant
total spectrum is clearly not thermal in the high energies.

\subsection{The Light Curve}

To construct the observed light curve $\varepsilon(t)$ we again
decompose the spherical plasma into concentric shells and consider two
effects:  First, is the relative arrival time of the first light from
each shell: light from outer shells will be observed before light from
inner shells;  Second, is the shape of the light curve from a single
shell.

Emission from moving pair plasma is beamed along the direction of travel
within an angle $\theta \sim 1/\gamma$.  The surface of simultaneity
of a relativistically expanding spherical shell, as seen by an observer,
is an ellipsoid \citep{fmn96}.  The observer time of intersection of an
expanding ellipse with a fixed shell of radius R as a function of
$\theta$ (i.e. the time at which emission from this intersection
circle is received) is:
\begin{equation}
t = \frac{R }{ v}(1 - v \cos \theta) \cong \frac{R }{ 2 \gamma^2 c}~~,
\label{ltcurvetimescale}
\end{equation}
for $\theta \ll 1, \gamma \gg 1$.  Integrating our boosted Planck
distribution of photons (Equation \ref{jay:E:planck}) over frequency,
we find that a relativistically expanding shell of radius R will have
a time profile (energy/time/steradian)
\begin{equation}
\varepsilon(\tau,v,T',R) = \frac{a }{ 2} \biggl(\frac{T' }{ \gamma \tau}
\biggr)^4 c~ R~ dR  \sim 1/\tau^4~~,
\label{jay:E:ltcurve}
\end{equation}
for $\tau > 1$ and where $\tau \equiv \frac{vt }{ R}$.  Emission
starts at $\tau_{i} = (1-v/c)$ and ends at $\tau_{f} = (1 + v/c)$.
The final light curve is constructed by summing the signal from all
shells.  The total thickness of the expanding plasma is $\sim c J/
\dot{J}$ because it expands at near the speed of light and $J/\dot{J}$
is the timescale of compression and coalescence which sets the
emission timescale.  Typically $R \sim 10^{12}$ cm and $J/\dot{J}
\sim$ a few seconds, so $c J/\dot{J} \ll R$ and the emitting plasma is
a thin shell.  The duration of the burst is determined by the duration
of emission because the observed timescale of emission from the plasma
shell is very short, $R/2\gamma^2 c \sim 0.01$ seconds (Equation
\ref{ltcurvetimescale}) for $\gamma \sim 100$, compared to
$J/\dot{J}$.

\section{ Results of Pair Plasma Emission } \label{resultspplsma}

\begin{figure}
\centering \epsfig{figure=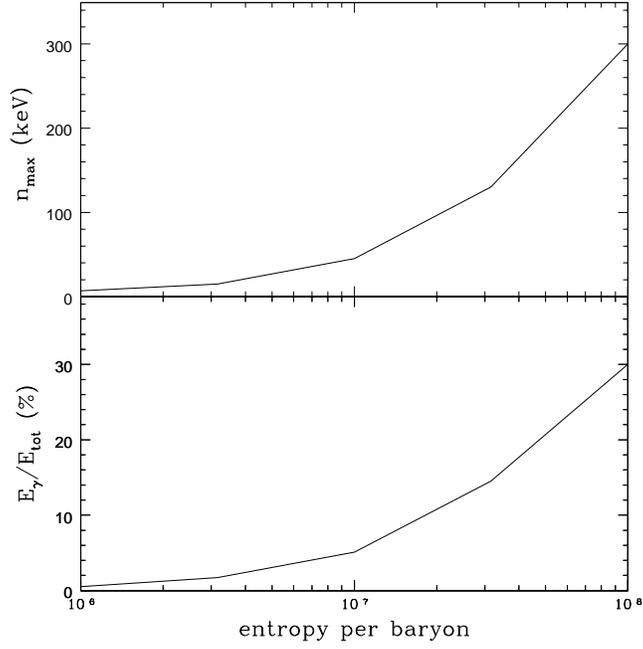, width=9cm}
\caption{The photon energy at the spectrum peak, and gamma-ray
efficiency are plotted for a total energy $E_{tot} = 10^{52}$ ergs
over a range of entropies $10^6$ to $10^8$. \label{date52}}
\end{figure}

\begin{figure}
\centering \epsfig{figure=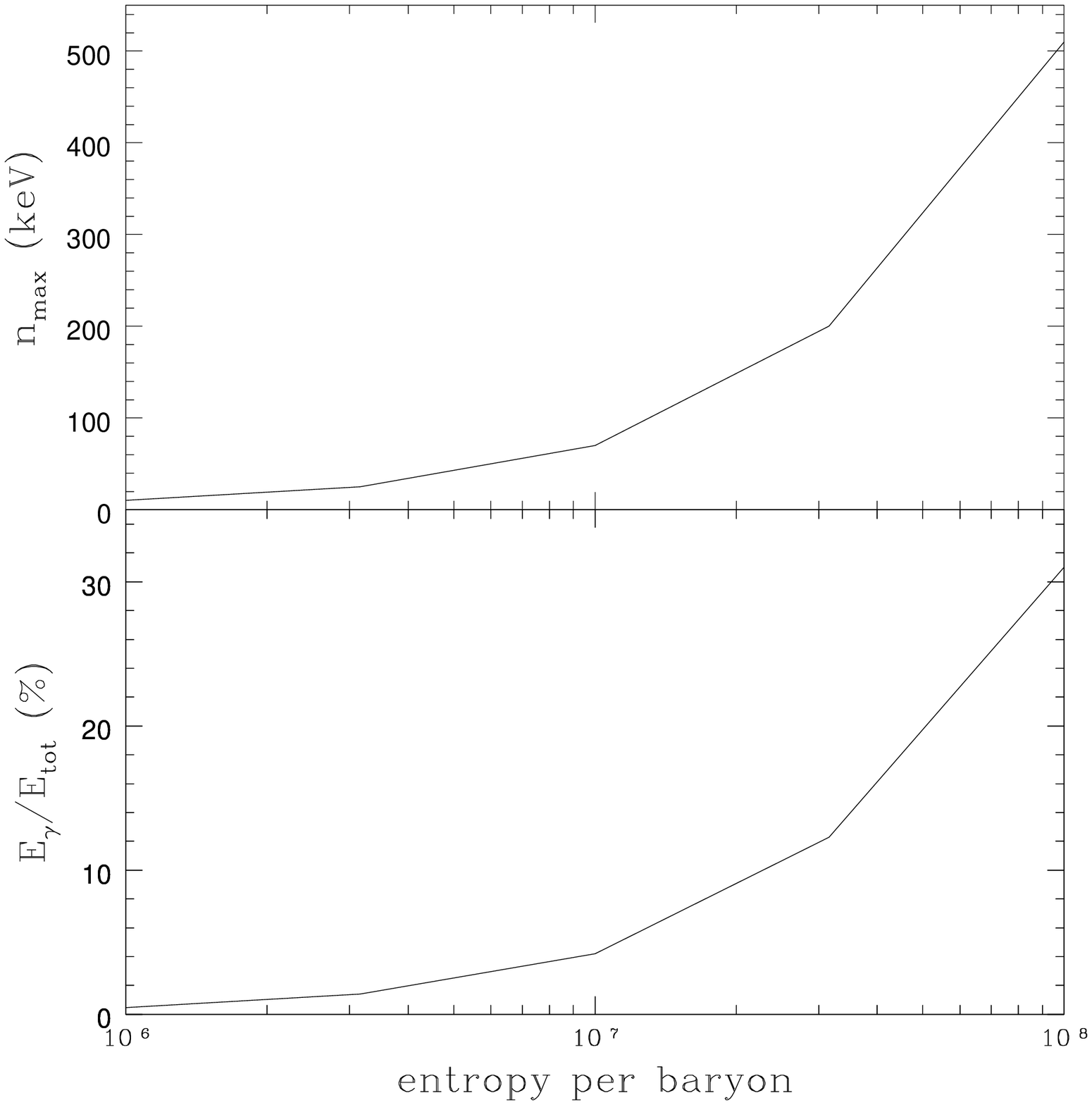, width=9cm} 
\caption{The photon energy at the spectrum peak, and gamma-ray
efficiency are plotted for a total energy $E_{tot} = 10^{53}$ ergs
over a range of entropies $10^6$ to $10^8$. \label{date53}}
\end{figure}

\begin{figure}
\centering \epsfig{figure=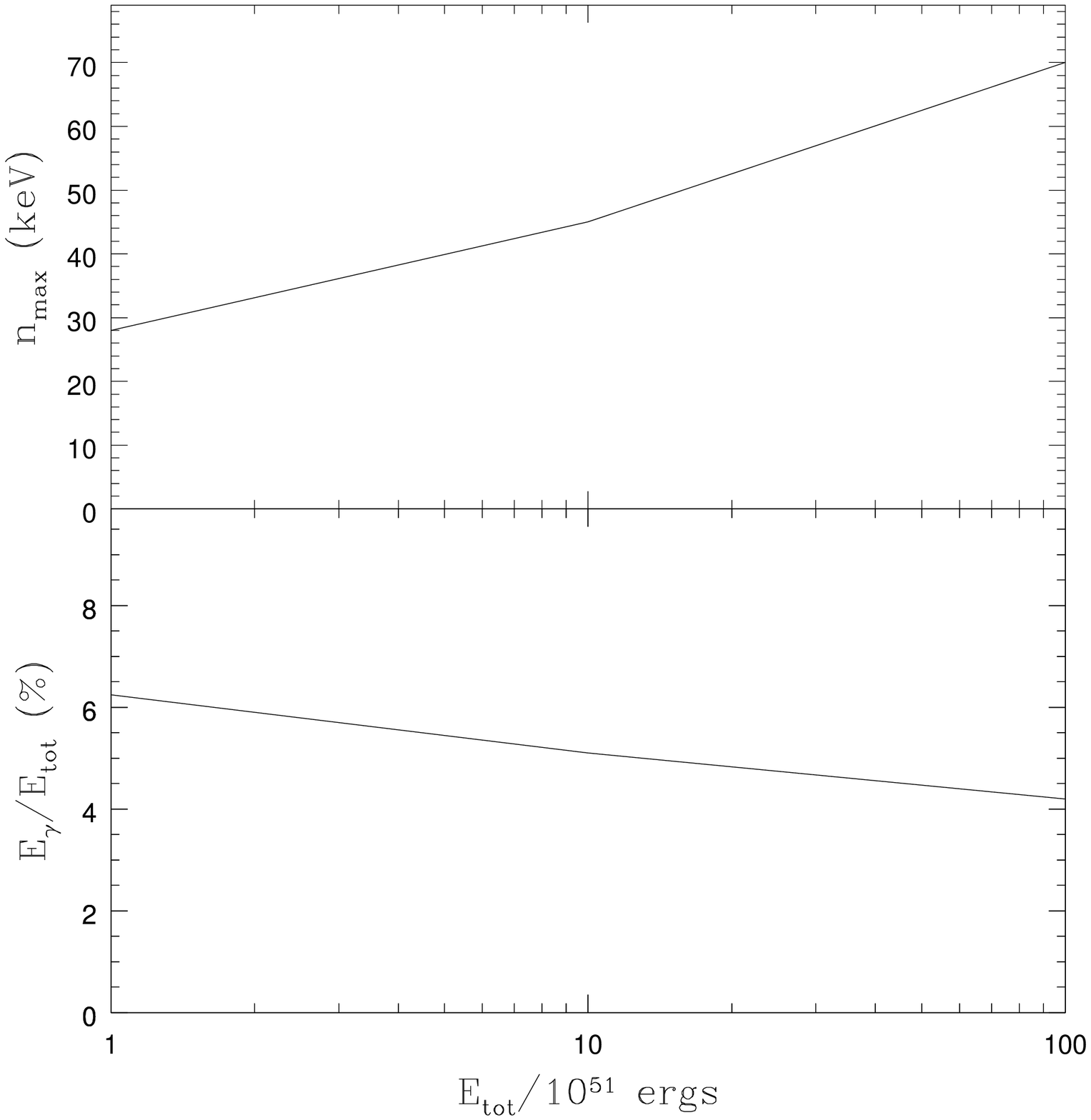, width=9cm}
\caption{The photon energy at the spectrum peak, and gamma-ray
efficiency are plotted for an entropy per baryon $s = 10^7$ over a
range of energies $10^{51}$ to $10^{53}$ ergs. \label{dats7}}
\end{figure}

We have run a variety of models over a range of entropies per baryon
and total energies.  The results are summarized in Figures
\ref{date52}, \ref{date53} \& \ref{dats7}.  We see that more powerful
bursts are derived from higher entropies per baryon and higher total
energies.  In particular, entropies per baryon of a few $\times 10^7$
allow a burst with a spectral peak $\sim 100$ keV and efficiencies
$E_\gamma/E_{tot} \sim 10$\%.  This is consistent with, although at
the upper end of the range of, the entropies calculated for the
$e^+e^-$ plasma deposited above the neutron stars.  Much further work
needs to be done to better characterize the nature of the stellar
compression and energy transport within the stars.  Also, more
elaborate simulations must be done to resolve the plasma flow in three
dimensions and to consider the effects of magnetic fields.  In Table
\ref{gammatable} we see the final Lorentz factor for a range of
expanding fireballs.  This data will be used when we look at the
collision of the fireball into an external medium.

\section{ External Shock Emission }

In previous sections the emission from an expanding fireball was
studied.  We found that the resulting emission spectrum and total
energy strongly depends upon the energy of the plasma
deposited near the surface of the neutron stars; entropies of
$\lesssim 10^6$ resulted in weak emission with most of the original
energy manifesting itself as kinetic energy of the baryons.  Thus, for
the low entropy per baryon fireballs ($s \sim 10^5 - 10^6$) produced
by NSBs it is necessary to examine the emission due to the interaction
of the relativistically expanding baryon wind with the interstellar
medium (ISM).

After becoming optically thin and decoupling with the photons, the
matter component of the fireball continues to expand and interact with
the ISM via collisionless shocks.  As the ISM is swept up, the matter
decelerates.  We model this process as an inelastic collision between
the expanding fireball and the ISM as in, for example,
\citet{piran98}.  We assume that the absorbed internal energy is
immediately radiated away.  From this we construct a simple picture of
the emission due to the matter component of the fireball
``snowplowing'' into the ISM of baryon number density $n$.

For a shell of a given rest mass $M$ expanding at Lorentz factor
$\gamma$, the conservation of momentum leads to
the following constraint equation:
\begin{equation}
\frac{d\gamma }{ \gamma^2 - 1} = - \frac{dM }{ M} ~,
\end{equation}
which has the solution
\begin{equation}
\frac{M}{M_0} = \sqrt{\frac{(\gamma_0 - 1) (\gamma + 1)}{(\gamma_0 +
1) (\gamma - 1)}} ~.
\end{equation}
Now we can put this in terms of radius by noting
\begin{equation}
M = M_0 + \frac{4 \pi}{3} n m_p c^2 R^3 ~.
\end{equation}
Thus,
\begin{equation}
R(\gamma) = R_0 \biggl( \frac{M}{M_0} - 1 \biggr)^{1/3} \cong R_0
\biggl(\frac{1}{\gamma} - \frac{1}{\gamma_0} \biggr)^{1/3} \quad
\text{for $\gamma$, $\gamma_0 \gg 1$} \label{Rgamma}~~,
\end{equation}
where 
\begin{equation}
R_0 \equiv \sqrt[3]{\frac{3 M_0}{ 4 \pi n m_p c^2}}~~,
\end{equation}
is the radius at which $M = 2 M_0$.  This is the characteristic radius
at which the shock decelerates.

We assume that the local thermal energy radiated away after a thin
shell of ISM mass $dM$ is swept up by the shock is
\begin{equation}
dE' = (\gamma - 1) dM ~. \label{dE}
\end{equation}
The observer time elapsed for the mass to expand a distance $dR$ is
\begin{equation}
dt_{obs} = \frac{ dR }{ 2 \gamma^2 c } ~. \label{dtobs}
\end{equation}
Equations (\ref{Rgamma},\ref{dtobs}) can be solved in the relativistic limit
to give
\begin{equation}
t(\gamma,\gamma_0) \cong \frac{R_0}{28 c} \biggl(\frac{9}{\gamma_0^2}
+ \frac{3}{\gamma \gamma_0} + \frac{2}{\gamma^2} \biggr) \biggl(
\frac{1}{\gamma} - \frac{1}{\gamma_0} \biggr)^{1/3} \quad \text{for
$\gamma$, $\gamma_0 \gg 1$}~ . \label{tgamma}
\end{equation}
The implied observer luminosity, from Equations (\ref{dtobs}, \ref{dE}), is
\begin{equation}
L = \frac{dE }{ dt_{obs}} = \frac{\gamma dE' }{ dt_{obs}} \approx 8 \pi R^2
\gamma^4 n m_p c^3 \label{LumDE}
\end{equation}
for $\gamma \gg 1$.  Using Equation (\ref{tgamma}), a relativistic
($\gamma \gg 1$) solution for observed luminosity, in ergs sec$^{-1}$,
over several epochs is
\begin{equation}
L(t) \cong
  \begin{cases}
   2.68 \times 10^{50} n \gamma_{300}^8 t^2 ({1 - {6.27 \times 10^{-3}} 
        \gamma_{300}^8 n E^{-1}_{52} t^3})^{10/3}& t<t_1\\ \\
   7.88 \times 10^{51} n^{1/3} E_{52}^{2/3} \gamma^{8/3}_{300} 
        (0.32 \frac{t }{ t_{max}} - 0.15)^{2/3} 
        (1.15 - 0.32 \frac{t }{ t_{max}})^{10/3}& t_1<t<t_2\\ \\
   5.3 \times 10^{51} n \gamma_{300}^4 
        \bigl[\frac{E_{52} }{ n \gamma_{300}}\bigr]^{4/7} t^{2/7} 
        \bigl(1 - \sqrt{f(t)} \bigr)^4 
        {\bigl({f(t) + \sqrt{f(t)}} \bigr)^{2/3}}&  t>t_2
  \end{cases} \label{Lequation}
\end{equation}
where constant parameters are, $ E_{52} \equiv E/10^{52}$ ergs,
$\gamma_{300} \equiv \gamma_0/300$ and $n$ is in baryons cm$^{-3}$.
For $t > t_2$:
\begin{equation}
f(t) \equiv 1 - 1.05 \biggl( \frac{ t_{max} }{ t} \biggr)^{3/7} 
\end{equation}
and
\begin{equation}
t_{max} \equiv 3.5~ \sqrt[3]{\frac{ E_{52} }{ n \gamma_{300}^8 }}
        \qquad \text{seconds}
\end{equation}
is the observer time at maximum luminosity $L_{max}$:
\begin{equation}
L(t_{max}) = L_{max} = 1.3 \times 10^{51} n^{1/3} \gamma^{8/3}_{300} 
        E_{52}^{2/3}  \qquad \text{ergs/sec}~.
\end{equation}
The times at which the solutions for each epoch are spliced together
are roughly
\begin{align}
t_1 &\sim 0.6 t_{max} \\
t_2 &\sim 1.5 t_{max} ~.
\end{align}

\begin{figure}
\centering \epsfig{figure=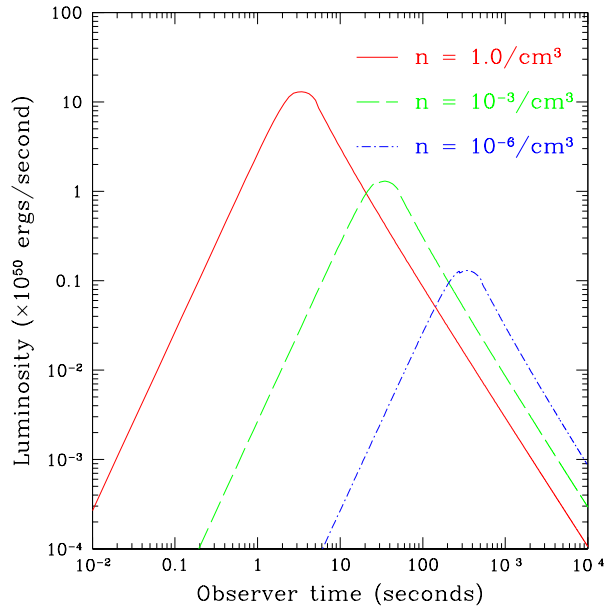, width=9cm}
\caption{The light curve for a $10^{52}$ erg fireball expanding at $\gamma =
300$ into interstellar media with three different baryon number
densities $n$. \label{xltcrv}}
\end{figure}

\begin{figure}[tb]
\centering \epsfig{figure=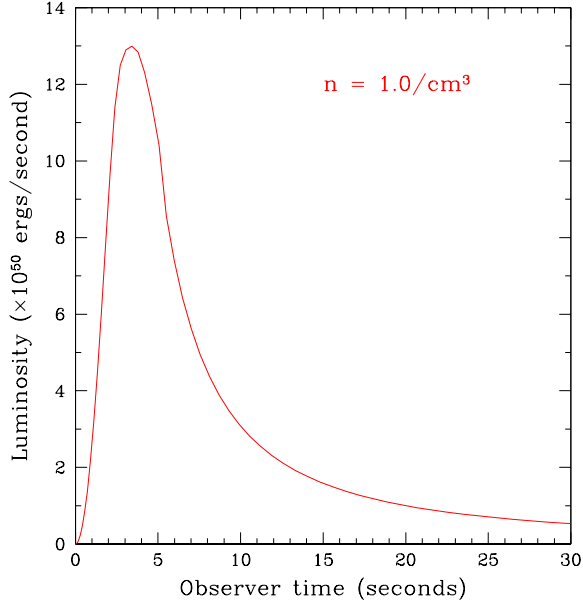, width=9cm}
\caption{The light curve for a $10^{52}$ erg fireball expanding at $\gamma =
300$ into interstellar media with baryon density $n = 1.0$
cm$^{-3}$. This curve is similar in its ``fast-rise, exponential
decay'' shape and $\sim 10$ second duration to the light curves of
many of the smooth-type GRBs.  \label{xltcrv2}}
\end{figure}

Figure \ref{xltcrv} shows the light curve for a $10^{52}$ erg fireball
expanding at $\gamma = 300$ for a range of ISM densities.  This
corresponds to an initial energy deposition above the neutron stars
with an entropy per baryon of $s = 10^5$ as seen in Table
\ref{gammatable}.  The expansion can be divided into a free-expansion
phase and a deceleration phase:
\begin{equation}
L(t) \propto
        \begin{cases}
        t^2& \text{free expansion phase     $(t < t_{max})$} \\
        t^{-10/7}& \text{deceleration phase     $(t > t_{max})$}
        \end{cases}
\end{equation}
Figure \ref{xltcrv2} shows a linear plot of the light curve for ISM
density $n = 1.0$ baryons cm$^{-3}$.  The ``fast-rise, exponential-decay''
or ``FRED''-like shape is evident and is in good qualitative agreement
with ``smooth'' GRBs.

\subsection{ Synchrotron Shock Spectrum }

Now we wish to model the spectrum of light emitted as the fireball
expands into the ISM.  To do this we assume an external synchrotron
shock model \citep{sm90,rm92,mr93b,rm94}.  This analysis is analogous to
afterglow models in the radiative limit.  Thus, the spectrum will have
the form \citep{spn98}
\begin{equation}
L_\nu =
        \begin{cases}
        (\nu/\nu_c)^{1/3} L_{\nu,max}&  \nu<\nu_c\\
        (\nu/\nu_c)^{-1/2} L_{\nu,max}& \nu_m > \nu>\nu_c\\
        (\nu_m/\nu_c)^{-1/2} (\nu/\nu_m)^{-p/2} L_{\nu,max}& \nu>\nu_m ~.
        \end{cases} \label{Lnu}
\end{equation}
Photon frequency $\nu_m$ is the frequency corresponding to the minimum
energy of the electron distribution above which the electrons are
assumed to have a power law functional form $n(\gamma) \sim
\gamma^{-p}$.  In the numerical examples that follow, we take 
the spectral index to be 
$p = 2.5$.  This is consistent with that calculated for
ultrarelativistic shocks \citep{bo98}.  The ``cooling frequency''
$\nu_c$ corresponds  to the energy below which the electrons cannot cool
on a hydrodynamic timescale.  The peak of the luminosity
spectrum is
\begin{equation}
        L_{\nu,max} \cong \biggl(\frac{p-2}{2 p - 2}\biggr)
        \frac{L}{\sqrt{\nu_m \nu_c}}~~,
\end{equation}
assuming $\nu_m \gg \nu_c$, which is valid throughout the burst
duration.  

There are two free parameters in this model.  $\epsilon_e$ is the
fraction of the kinetic energy of the baryons that is deposited into
the electrons by the shock.  $\epsilon_B$ is the ratio of the magnetic
field energy density to the kinetic energy density of the baryons.  In
these simulations we take each of these values to be $1/4$.

The evolution of the characteristic frequency $\nu_m$ is described by
\begin{equation}
        \nu_m \cong 1.4 \times 10^{4} \epsilon^2_e \epsilon_B^{1/2}
\sqrt{n} \gamma_{300}^4~ \text{keV}
        \begin{cases}
        (1 - \bigl(\frac{2 c t}{R_0}\bigr)^3 \gamma_0^7)^4&
t<t_1\\ \\
        (1.15 - 0.32 t/t_{max})^4& t_1<t<t_2\\ \\
        \biggl(1 - \sqrt{1 - \frac{2}{\gamma_0} (\frac{14 c
t}{R_0})^{-3/7}}\biggr)^4& t>t_2 ~.
        \end{cases} \label{numequation}
\end{equation}

\begin{figure}
\centering \epsfig{figure=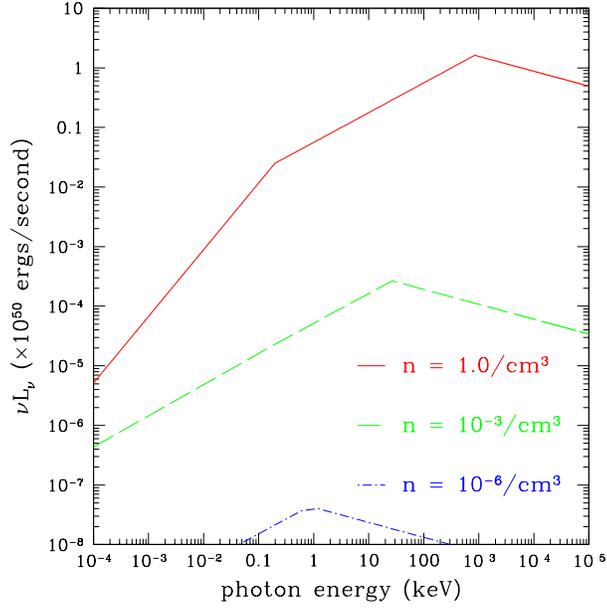, width=9cm}
\caption{The synchrotron spectrum for a $10^{52}$ erg fireball expanding at
$\gamma = 300$ into interstellar media with three different baryon
number densities $n$. \label{xspec}}
\end{figure}

\begin{figure}
\centering \epsfig{figure=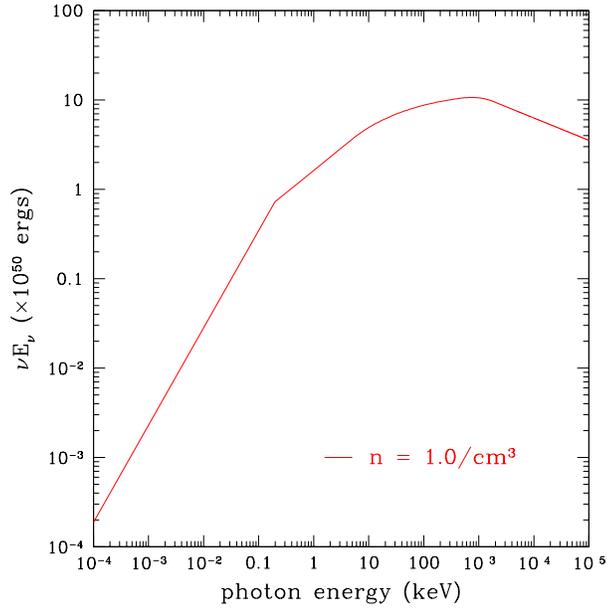, width=9cm}
\caption{The total energy spectrum of a $10^{52}$ erg fireball expanding at
$\gamma = 300$ into an interstellar medium with baryon number density
$n$ = 1.0 cm$^{-3}$. \label{xfluence}}
\end{figure}

The behavior of the cooling frequency $\nu_c$ is more difficult to
characterize since it depends on the hydrodynamical timescale of the
fluid. Fortunately, however, $\nu_c$ is much smaller than $\nu_m$.
Therefore, its exact behavior is not important for this analysis.  
Thus, we assume
$\nu_c$ to be constant at early times and follow its asymptotic
power-law at later times:
\begin{equation}
        \nu_c \cong 2.7 \times 10^{-3} \epsilon_B^{-3/2} E_{52}^{-4/7}
\gamma_{300}^{4/7} n^{-13/14}~ \text{keV}
        \begin{cases}
        t_{max}^{-2/7}&  t \leq t_{max}\\ 
        t^{-2/7}& t > t_{max}  ~~.
\end{cases} 
\label{nucequation}
\end{equation}  
The spectrum of the burst at peak luminosity $L_{max}$ is shown in
Figure \ref{xspec}.  For $n = 1.0$ baryons cm$^{-3}$, most of the energy is
emitted at photon energies $\sim 100$ keV.  Using Equations
(\ref{Lequation},\ref{numequation},\ref{nucequation}) for $L$, $\nu_m$
and $\nu_c$ respectively, we can determine the spectrum (Equation
\ref{Lnu}).  The fluence spectrum of the burst is obtained by integrating the
evolving luminosity spectrum (Equation \ref{Lnu}) in time.  This is shown in
Figure \ref{xfluence}.  This figure again shows that most of the burst energy is
in photons of several hundred keV energy.

Now we can ask what the efficiency is of gamma-ray production by the
shock compared to other wavelengths.  At any time, the fraction of
luminosity above a given minimum frequency $\nu_{min}$ is

\begin{equation}
\varepsilon_{ff} = 
        \begin{cases} 
        \frac{\bigl[ \bigl( \frac{2 p -
                2}{p-2} \bigr) \nu_m^{-1/2} - 2 \nu_{min}^{1/2}
                \bigr]}{\bigl[ \bigl(\frac{2 p - 2}{p-2} \bigr)
                \nu_m^{-1/2} - \frac{5}{4} \nu_c^{1/2} \bigr]}&
                \nu_c < \nu_{min} < \nu_m\\ \\
        \frac{ \bigl(\frac{2}{p - 2} \bigr)
                \biggl(\frac{\nu_m^{p-1}}{\nu_{min}^{p-2}} \biggr)^{1/2}}
                {\bigl[ \bigl(\frac{2 p - 2}{p-2} \bigr)
                \nu_m^{-1/2} - \frac{5}{4} \nu_c^{1/2} \bigr]}&
                \nu_{min} > \nu_c ~.
        \end{cases}
\end{equation}
Thus, we can calculate the duration, $t_{90}$, of the luminosity at
energies above this minimum energy.  This is done in Table
\ref{t90table} for the various fireballs shown in Table
\ref{gammatable}.  There is a competition between factors limiting the
 duration; lower energy fireballs simply have fewer high energy photons,
and thus, shorter duration; while higher energy fireballs expand and
evolve faster and thus have shorter duration.  Fireballs with energy
of order $10^{52}$ ergs and entropies per baryon of order $10^5$ yield
a value for $t_{90}$ which is consistent with observation.

The overall efficiency of the production of photons above a frequency
$\nu_{min}$ is 
\begin{equation}
\begin{split}
\varepsilon_{ff tot} &\equiv \frac{ \int_0^\infty
        \int_{\nu_{min}}^\infty L_\nu d \nu dt } { \int_0^\infty
        \int_0^\infty L_\nu d \nu dt } \\ &\approx 1 -
        \biggl(\frac{\nu_{min}}{\nu_{max}} \biggr)^{1/6} \quad
        \text{for  } \nu_{min} \ll \nu_{max}~~,
\end{split}
\end{equation}
where $\nu_{max}$ is the value of $\nu_m$ (Equation \ref{numequation})
at $t = t_{max}$.  For $\nu_{min} = 10$ keV we have an overall
efficiency of about $\varepsilon_{ff tot} \approx 75$ \%.  Thus, the
radiative external shock GRB is quite efficient at producing
gamma-rays if our assumptions are reasonable.

\section{Conclusions}

In this paper we have argued that  heated neutron stars
(perhaps by compression of close 
neutron-star binaries) are 
viable candidates  for the production of large, high entropy per baryon,
$e^+e^-$ pair plasma fireballs, and thus, for the creation of gamma-ray
bursts.  We find that fireballs of total energy $E \sim 10^{51}$ to 
$3 \times 10^{52}$ ergs and an entropy per baryon of $s/k \sim 10^5 - 10^6$ are
possible.  Values for the  entropy as high as $10^7$ may be realized
during the peak $\nu
\overline{\nu}$ luminosity (of $\approx 10^{53}$ ergs sec$^{-1}$).
Emergent gamma-rays yield
a quasi-thermal spectrum peaked at $\sim 100$ keV with an efficiency
of conversion from pair plasma to photons of $\sim 30$\%.  
The lower entropy component of the fireball will
initiate a shock which propagates into the ISM, generating an external
shock GRB.

The calculation utilizing the supernova computer program (Section
\ref{neutrinoannihilation}) to describe the neutrino and matter transport,
produces a baryonic wind that 
contains $\sim 90$\% neutrons. The decay of the neutrons to protons occurs on the same
time scale as that for which the protons are decelerated by the
intersteller medium. This delayed conversion of neutrons to protons
will broaden the gamma-ray signal by a factor of a few. In
addition, the decay electrons will strongly increase the entropy of the
expanding plasma at the late times. In future work we  will quantify the role
of neutrons and explore the possibility of fireball photons
inverse-Compton up-scattering off of the accelerated electron
distribution of the external shock. This corresponds to the emission scenario
put forward by \citet{lksc97}.

As of yet this model is spherically symmetric. Thus, it  can only generate
bursts with a smooth light-curve structure.  However, we expect a large variety
of GRB morphologies with varied time structure due to: 1) three
dimensional resolution of the plasma flow; 2) plasma instabilities due
to increased heating of the deposited plasma with time; and 3)
variation in the ratio of star mass in the NSB, effecting the relative
compression and heating rate of each star.

In future work we will numerically model, in three dimensions, the flow
of the $e^+e^-$ pair plasma in the midst of the orbiting neutron
stars.  We have written a three-dimensional general relativistic
hydrodynamic code to study this three-dimensional behavior.  In
particular, we wish to study the possible formation of jets along the
orbital axis due to the collision of plasma blowing away
from each star.  Also, we have done simulations which suggest that the
internal magnetic field of the neutron stars may be high.
Thus, the inclusion of 
magnetohydrodynamic plasma effects including Alfv\'en instabilities
and reconnections may ultimately be necessary.

The authors wish to thank the late Jean-Alain Marck for his inciteful and
encouraging remarks on an earlier version of this manuscript.  
This work was performed under the auspices of the U.S. Department of
Energy by University of California Lawrence Livermore National
Laboratory under contract W-7405-ENG-48.  J.R.W. was partly supported
by NSF grant PHY-9401636.  Work at University of Notre Dame supported
in part by DOE grant DE-FG02-95ER40934, NSF grant PHY-97-22086, and by
NASA CGRO grant NAG5-3818.  \\



\appendix
\section{Appendix: Neutron Star EOS}

 A key requirement of a gamma-ray burst paradigm based upon collapsing
neutron stars in binaries is that the equation of state be relatively
soft so that significant compression and heating can occur before
inspiral.  Therefore, for completeness in this Appendix we review
arguments for and against a ``soft'' neutron star EOS.

The neutron star EOS must
extend from normal iron nuclei on the surface to as much as 15 times
nuclear matter density in the interior.  At the same time, one must
consider that  neutron stars in weak-interaction equilibrium are highly
isospin asymmetric.  They may also carry net strangeness.  Therefore,
only pieces of the neutron-star equation of state, e.g. the nuclear
compressibility, are accessible in laboratory experiments.  
The value for the nuclear
compressibility $K_s$ can be derived from the nuclear monopole
resonance \citep{blais80}.  The present value ($K_s = $ 230 MeV) is
consistent with a modestly soft nuclear equation of state.  

Nuclear heavy-ion collision data can also be used to shed some insight, particularly for
the heated neutron-star equation of state.  For example,
\citet{mw94b} studied heavy ion
collisions of $^{139}$La on $^{139}$La as a means to constrain the
supernova EOS.  The electron fraction for $^{139}_{57}$La ($Y_e = 0.41$)
overlaps that of supernovae which range from $Y_e = 0.05$ to 0.50.
They showed that the pion contribution to the EOS could be constrained
by the observed pion multiplicities from central collisions.  The formation and
evolution of pions was computed in the context of Landau-Migdal theory
to model the effective energy and momenta of the pions.  A key
aspect of hydrodynamic simulations of the heavy-ion data was the
determination of the Landau parameter $g'$.  Their determination of
the pion contribution to the equation of state implies a 
relatively soft
equation of state after pion condensation such that
a maximum neutron-star mass of $M \le 1.64$
M$_\odot$ is inferred.

There have been dozens of nuclear equations of state introduced over
the years.  Summaries of some of them can be found in
\citet{schaabw99} and \citet{ab77}.  As far as the maximum mass of a
neutron star is concerned, most theoretical equations of state fall
into two groups, those which only describe the mean nuclear field even
at high density and those which allow for various condensates,
e.g. pions, kaons, hyperons, and even quark-gluon plasma.  Table
\ref{tableos} summarizes the basic neutron star properties based upon
most available nuclear equations of state \citep{lattimer98}.

Equations of state which are based upon the mean nuclear field tend to
be ``stiff'' at high density.  Therefore, they reach lower interior
densities for the same baryonic mass and tend to allow a higher
maximum neutron-star mass $m_{max} \sim 1.8-2.2$ M$_\odot$.  Such
equations of state also tend to become acausal at the high densities
associated with the maximum neutron-star mass.  On the other hand, the
relativistic equations of state are generally causal at high density.
They also tend to be somewhat ``soft'', therefore allowing a higher
central density for a given baryon mass and generally implying a
maximum neutron-star mass in the range $m_{max} \sim 1.3-1.7 $
M$_\odot$.  We note, however, that recent 3-body corrections to a
relativistic EOS \citep{pandharipande99} tend to stiffen an otherwise
soft relativistic EOS.

For the most part, constraints on the neutron star equation of state
must ultimately come from observations of neutron stars themselves.  Over the
years attempts have been made with limited success to constrain the
equation of state based upon the maximum observed rotation frequency
\citep[e.g.][]{fip86} or the thermal response to neutron star glitches
\citep[e.g.][]{page98}.  In recent years, however, new observational
constraints on the structure and properties of neutron stars are
becoming available \citep{lattimer98}.  Observations of quasi-periodic
oscillations (QPOs) \citep{szss+96,vszj+96,vwhc97}, pulsar light
curves \citep{yhh94,tc99}, and glitches \citep{lel99}, studies of
soft-gamma repeaters \citep{k+98,gvd99}; and even the identification
of an isolated non-pulsing neutron stars \citep{wwn96,h+97} have all
led to the hope that significant constraints on the mass-radius relation and maximum
mass of neutron stars may be soon coming.

\subsection{Pulsars}

Two possible constraints come from measured pulsar systems.  The most
precisely measured property of any pulsar system is its spin
frequency.  The frequencies of the fastest pulsars (PSR B1937+21 at
641.9 Hz and B1957+20 at 622.1 Hz) already constrain the equation of
state under the assumption that these pulsars are near their maximum
spin frequency \citep{fip86}.  In particular, the equation of
state cannot be too stiff, though maximum masses as large as 3
M$_\odot$ are still allowed.

A much more stringent constraint may come from the numerous
determinations of neutron-star masses in pulsar binaries.  There are
now about 50 known pulsars in binary systems.  Of these 50,
approximately 15 of them have significantly constrained masses.  These
are summarized in Table \ref{tablnsmass}.  The measured masses are all
consistent with low neutron-star masses in the range $m \approx 1.35
\pm 0.10$ M$_\odot$ \citep{tc99}.  Even though these masses are low,
this does not necessarily mean that the maximum neutron-star
mass is in this range.  If one adopts these masses as approaching
the maximum neutron-star mass, then the softer equations of state are
preferred.  However, this narrow mass range may be the result of the
mechanism of neutron-star formation in supernovae and not an
indication of the maximum neutron-star mass.

  In a recent paper, \citet{lel99} have proposed that glitches
observed in the Vela pulsar and six other pulsars may place some
constraint on the nuclear EOS.  In particular, if the glitches
originate from the liquid of the inner crust, and if the mass of the
Vela pulsar is 1.35 consistent with Table 2, then the radius of the
Vela pulsar must be $R ^>_\sim 8.9$ km.  This result is consistent
with either a soft or stiff equations of state.  A better theoretical
determination of the pressure at the crust-core interface might lead
to a more stringent constraint.

\subsection{QPO's}
The identification of kilohertz QPO's with the last stable orbit
around a neutron star also could significantly constrain the
neutron-star equation of state \citep[e.g.][]{schaabw99}.  For example,
demanding that the 1.2 khz QPO from source KS 1731-260 be the last
stable orbit requires a neutron-star mass of 1.8 M$_\odot$, On the
other hand, other interpretations are possible for the origin of QPO's.
For example, they could be a harmonic of a lower frequency outer
orbit, or they might result from effects closer to the neutron-star
surface.  Among proposals for the source of the QPO phenomenon are:
boundary layer oscillations \citep{chv98}; radial oscillations and
diffusive propagation in the transition region between the neutron
star and the last Keplerian orbit \citep{to99}; Lense-Thirring
precession for fluid particles near the last stable orbit
\citep{mlp98,mvsb99,ms99}; and nonequitorial resonant oscillations of
magnetic fluid blobs \citep{vs98}.

\subsection{Supernova constraints}

The lack of a radio pulsar in SN1987A, along with nucleosynthesis
constraints on the observed change of helium abundance with
metallicity has led to the suggestion \citep{bb94,bb95} that the
maximum neutron-star mass must be $_\sim^< 1.56$ M$_\odot$.  In this
picture, the development of a kaon condensate tends to greatly soften
the EOS after $\sim 12$ sec.  Thus, even though neutrinos were
emitted, the core subsequently collapses to a black hole.

One constraint comes from the
neutrino signal itself observed to arise from supernova SN1987A.
The fact that the neutrinos arrived over an interval of at least
twelve seconds implies a significant cooling and neutrino diffusion
time from the core.  This favors a soft equation of state in which
the core is more compact and at higher temperature in the supernova
models.  For example, the simulations of \citet{wm93} require
a maximum neutron-star mass of  $^<_\sim 1.6$ M$_\odot$.

\subsection{Isolated Neutron Star}
A most promising constraint on the neutron-star EOS may come from the
determination of the radius for the isolated nonpulsing neutron star
RX J185635-3754, first detected by ROSAT \citep{wwn96}.  The inferred
(redshifted) surface temperature from the X-ray emission is about 35
eV.  Atmospheric models of this emission then imply \citep{lattimer98,
alp98, wlr+99} that for a distance between 31 and 41 pc, a radius
between $5.75 < R/{\rm km} < 11.4$ and a mass of $1.3 < M < 1.8$, is
most consistent with the observed emission.  This is suggestive of a
soft equation of state.  However, this constraint requires that the
distance be less than 41 kpc.  On the other hand, \citet{wlr+99} find
that the cooling properties of the soft X-ray source RX J0720.4-3125
are most consistent with a moderately stiff or stiff EOS provided that
the age of this star is less than 10$^5$ yr.  Proper motion studies
with HST are currently underway to determine a reliable distance to RX
J185635-3754.  These studies will provide a key constraint on the
nuclear equation of state.


\begin{deluxetable}{ccc|ccc}
\tablecolumns{6} 
\tablewidth{0pc} 

\tablecaption{Table of the equation of state for a neutron star with
critical mass $M_c = 1.575 M_\odot$.  Values are baryonic
density $\rho$, specific energy $\epsilon$ and $\Gamma \equiv 1 +
P/\rho \epsilon$ where $P$ is the pressure. \label{eostable}}

\tablehead{\colhead{$\rho$ gm/cm$^{-3}$} & \colhead{$\epsilon$ ergs/gm} & \colhead{$\Gamma$} & \colhead{$\rho$ gm/cm$^{-3}$} & \colhead{$\epsilon$ ergs/gm} & \colhead{$\Gamma$} }
\startdata
1.00 $\times 10^{9}$         &  $1.11 \times 10^{18}$        &   1.386 & $1.46 \times 10^{13}$        &  $1.94 \times 10^{19}$         &   1.150 \\
$1.46 \times 10^{9}$         &  $1.28 \times 10^{18}$         &   1.380 & $2.15 \times 10^{13}$        &  $2.03 \times 10^{19}$         &   1.108  \\
$2.15 \times 10^{9}$         &  $1.47 \times 10^{18}$         &   1.372 & $3.16 \times 10^{13}$        &  $2.09 \times 10^{19}$         &   1.057 \\
$3.16 \times 10^{9}$         &  $1.69 \times 10^{18}$         &   1.367 & $4.64 \times 10^{13}$        &  $2.13 \times 10^{19}$         &   1.052 \\
$4.64 \times 10^{9}$         &  $1.94 \times 10^{18}$         &   1.363 & $6.81 \times 10^{13}$        &  $2.17 \times 10^{19}$         &   1.061 \\
$6.81 \times 10^{9}$         &  $2.22 \times 10^{18}$         &   1.358 & $1.00 \times 10^{14}$        &  $2.23 \times 10^{19}$         &   1.093 \\
$1.00 \times 10^{10}$        &  $2.54 \times 10^{18}$         &   1.352 & $1.46 \times 10^{14}$        &  $2.36 \times 10^{19}$         &   1.213 \\
$1.46 \times 10^{10}$        &  $2.90 \times 10^{18}$         &   1.346 & $2.15 \times 10^{14}$        &  $2.68 \times 10^{19}$         &   1.468 \\
$2.15 \times 10^{10}$        &  $3.30 \times 10^{18}$         &   1.341 & $3.16 \times 10^{14}$        &  $3.39 \times 10^{19}$         &   1.778 \\
$3.16 \times 10^{10}$        &  $3.75 \times 10^{18}$         &   1.336 & $4.64 \times 10^{14}$        &  $4.75 \times 10^{19}$         &   1.992 \\
$4.64 \times 10^{10}$        &  $4.26 \times 10^{18}$         &   1.330 & $6.81 \times 10^{14}$        &  $7.09 \times 10^{19}$         &   2.103 \\
$6.81 \times 10^{10}$        &  $4.82 \times 10^{18}$         &   1.322 & $1.00 \times 10^{15}$        &  $1.09 \times 10^{20}$         &   2.16 \\
$1.00 \times 10^{11}$        &  $5.44 \times 10^{18}$         &   1.314 & $1.46 \times 10^{15}$        &  $1.69 \times 10^{20}$         &   2.145 \\
$1.46 \times 10^{11}$        &  $6.13 \times 10^{18}$         &   1.307 & $2.15 \times 10^{15}$        &  $2.58 \times 10^{20}$         &   2.063 \\
$2.15 \times 10^{11}$        &  $6.88 \times 10^{18}$         &   1.300 & $3.16 \times 10^{15}$        &  $3.87 \times 10^{20}$         &   2.061 \\
$3.16 \times 10^{11}$        &  $7.71 \times 10^{18}$         &   1.294 & $4.64 \times 10^{15}$        &  $5.81 \times 10^{20}$         &   2.059 \\
$4.64 \times 10^{11}$        &  $8.62 \times 10^{18}$         &   1.288 & $6.81 \times 10^{15}$        &  $8.67 \times 10^{20}$         &   2.03 \\
$6.81 \times 10^{11}$        &  $9.61 \times 10^{18}$         &   1.280 & $1.00 \times 10^{16}$        &  $1.28 \times 10^{21}$         &   2.015 \\
$1.00 \times 10^{12}$        &  $1.06 \times 10^{19}$         &   1.270 & $1.46 \times 10^{16}$        &  $1.88 \times 10^{21}$         &   2.007 \\
$1.46 \times 10^{12}$        &  $1.17 \times 10^{19}$         &   1.261 & $2.15 \times 10^{16}$        &  $2.76 \times 10^{21}$         &   2.003 \\
$2.15 \times 10^{12}$        &  $1.29 \times 10^{19}$         &   1.250 & $3.16 \times 10^{16}$        &  $4.05 \times 10^{21}$         &   2.001 \\
$3.16 \times 10^{12}$        &  $1.41 \times 10^{19}$         &   1.236 & $4.64 \times 10^{16}$        &  $5.94 \times 10^{21}$         &   2.001 \\ 
$4.64 \times 10^{12}$        &  $1.54 \times 10^{19}$         &   1.224 & $6.81 \times 10^{16}$        &  $8.71 \times 10^{21}$         &   2 \\
$6.81 \times 10^{12}$        &  $1.67 \times 10^{19}$         &   1.216 & $1.00 \times 10^{17}$        &  $1.27 \times 10^{22}$         &   2 \\
$1.00 \times 10^{13}$        &  $1.81 \times 10^{19}$         &   1.211  \\
\enddata
\end{deluxetable}

\begin{deluxetable}{ccc}
\tablecolumns{3} 
\tablewidth{0pc} 
\tablecaption{Central density and released gravitational energy 
as a function of the binary angular momentum $J$ in geometrized units.
This calculation \citep{mw99} is for
a neutron star with M$_B = 1.548$ M$_\odot$, M$_G = 1.39$ M$_\odot$,
 and and EOS for
which M$_c = 1.575$ M$_\odot$.}
\tablehead{\colhead{J ($10^{11}$ cm$^{2}$)} & \colhead{$\rho$ ($10^{15}$ g cm$^{-3}$)} & \colhead{$E$ ($10^{52}$ erg)}} 
\startdata
1.65 & 1.48 & 4.3\\
1.80 & 1.46 & 2.4\\
2.0  & 1.45 & 1.5\\
2.2  & 1.43 & 1.0\\
2.4  & 1.40 & 0.7\\
2.6  & 1.38 & 0.6\\
$\infty$  & 1.34 & 0\\
\enddata
\label{tableheat}
\end{deluxetable}

\begin{deluxetable}{cccc}
\tablecolumns{4} 
\tablewidth{0pc} 
\tablecaption{Final Lorentz factor of the baryon wind for a range
of initial total energies and entropies per
baryon. \label{gammatable}}

\tablehead{
\colhead{} &\multicolumn{3}{c}{Entropy per Baryon} \\\cline{2-4}
\colhead{Energy (ergs)} & \colhead{$10^5$} & \colhead{$10^6$} & \colhead{$10^7$}}
\startdata
$10^{51}$ &175 &1750 & $1.6 \times 10^4$ \\
$10^{52}$ &350 &3100 & $2.9 \times 10^4$ \\
$10^{53}$ &525 &5500 & $5.3 \times 10^4$ \\
\enddata
\end{deluxetable}

\begin{deluxetable}{cccc}
\tablecolumns{4} 
\tablewidth{0pc} 
\tablecaption{ The $t_{90}$ duration, in seconds, of external shock
GRBs corresponding to the fireballs outlined in Table
\ref{gammatable}.  This $t_{90}$ is calculated energy emitted in photons
greater than 10 keV. \label{t90table}}

\tablehead{
\colhead{} &\multicolumn{3}{c}{Entropy per Baryon} \\\cline{2-4}
\colhead{Energy (ergs)} & \colhead{$10^5$} & \colhead{$10^6$} & \colhead{$10^7$}}
\startdata
$10^{51}$ &40.3 &0.7 & $3.7 \times 10^{-3}$ \\
$10^{52}$ &21.6 &7.2 & $1.6 \times 10^{-3}$ \\
$10^{53}$ &23.8 &0.17 & $1.8 \times 10^{-3}$ \\
\enddata
\end{deluxetable}

\begin{center}
\begin{table}
\caption{Neutron star properties from various equations of state}
\begin{tabular}{lccc}
\tableline 
\tableline \\
Equation of State & Composition & Maximum Mass (M$_\odot$)& R (km)\\
\tableline \\
Mean Nuclear Field& $p,n,e^-, \mu^-$ & $\approx  2.0 \pm 0.20 $  & $ \approx 13 \pm 3$\\

\tableline \\
Exotic Particles/ & $p,n,e^-, \mu^-, \Lambda, \Sigma^{\pm,0},
\Xi^{0,-}$, &  $\approx 1.5 \pm 0.2 $ & $\approx 9 \pm 1$ \\
Condensates & $ \Delta^{\pm,0.++}, K^{\pm,0}, \pi^{\pm,0}$, quarks, etc.  \\

\tableline \\
\end{tabular}
\label{tableos}
\end{table}
\end{center}

\begin{table}
\caption{Summary of masses of observed pulsars in binaries.} 
\begin{tabular}{lccc}
\tableline 
\tableline \\
Pulsar & Mass (M$_\odot$)& &\\
\tableline \\
{\it Double Neutron Star Systems} \\
\tableline \\
J1518+4904 & 1.56 $^{0.13}_{0.44}$ \\
J1518+4904 & 1.05 $^{0.45}_{0.11}$ \\
B1534+12   & 1.339 $\pm 0.0003$ \\
B1534+12   & 1.339 $\pm 0.0003$ \\
B1913+16   & 1.4411 $\pm 0.00035 $    \\
B1913+16   & 1.3874 $\pm 0.00035$     \\
B2127+11C  & 1.349 $\pm 0.040$       \\
B2127+11C  & 1.363 $\pm 0.040$       \\
B2303+46   & 1.30 $^{+0.13}_{-0.46}$ \\
B2303+46   & 1.34 $^{+0.47}_{-0.13}$ \\
\tableline \\
{\it Neutron Star/White-Dwarf Systems} \\
\tableline \\
J1012+5307 & 1.7 $\pm 0.5$  \\
J1713+0747 & 1.45 $\pm 0.31$ \\
J1713+0747 & 1.34 $\pm 0.20$ \\
B1802-07   & 1.26 $^{+0.08}_{-0.17}$ \\
B1855+09   & 1.41 $\pm 0.10$ \\
\tableline \\
{\it Neutron Star/Main-Sequence Systems} \\
\tableline \\
J0045-7319 & 1.58 $\pm 0.34$ \\
\tableline \\
\end{tabular}
\label{tablnsmass}
\end{table}


\end{document}